\documentclass[manuscript]{emulateapj}
\usepackage{graphicx}
\usepackage{amssymb}
\usepackage{epstopdf}
\usepackage{ctable}
\newcommand{\degree}{\mbox{$^{\circ}$}}
\usepackage{multirow}
\usepackage{longtable}
\usepackage{threeparttable}
\usepackage{natbib}
\begin{document}

\title{HST Imaging of Dust Structures and Stars in the Ram Pressure Stripped Virgo Spirals NGC 4402 and NGC 4522: Stripped from the Outside In with Dense Cloud Decoupling}
\author{A. Abramson, J. Kenney, H. Crowl, T. Tal}


\begin{abstract}

We describe and constrain the origins of ISM structures likely created by ongoing ICM ram pressure stripping in two Virgo Cluster spirals, NGC 4522 and NGC 4402, using HST BVI images of dust extinction and stars, as well as supplementary HI, H$\alpha$, and radio continuum images. With a spatial resolution of $\sim$10 pc in the HST images, this is the highest-resolution study to date of the physical processes that occur during an ICM-ISM ram pressure stripping interaction, ram pressure stripping's effects on the multi-phase, multi-density ISM, and the formation and evolution of ram-pressure-stripped tails. In dust extinction, we view the leading side of NGC 4402 and the trailing side of NGC 4522, so we see distinct types of features in both. In both galaxies, we identify some regions where dense clouds are decoupling or have decoupled and others where it appears that ~kpc-sized sections of the ISM are moving coherently. NGC 4522 has experienced stronger, more recent pressure and has the ``jellyfish" morphology characteristic of some ram pressure stripped galaxies. Its stripped tail extends up from the disk plane in continuous upturns of dust and stars curving up to $\sim$2 kpc above the disk plane. On the other side of the galaxy, there is a kinematically and morphologically distinct extraplanar arm of young, blue stars and ISM above a mostly-stripped portion of the disk, and between it and the disk plane are decoupled dust clouds that have not been completely stripped. The leading side of NGC 4402 contains two kpc-scale linear dust filaments with complex substructure that have partially decoupled from the surrounding ISM. NGC 4402 also contains long dust ridges, suggesting that large parts of the ISM are being pushed out at once. Both galaxies contain long ridges of polarized radio continuum emission indicating the presence of large-scale ordered magnetic fields. We propose that magnetic fields could bind together gas of different densities, causing nearby gas of different densities to be stripped at the same rate and creating the large, coherent dust ridges and upturns. A number of factors likely play roles in determining what types of structures form as a result of ram pressure, including ram pressure strength and history, the location within the galaxy relative to the leading side, and pre-existing substructure in the ISM that may be bound together by magnetic fields during stripping.

\end{abstract}

\section{Introduction}

The evolution of galaxies in clusters is strongly affected by their environment \citep[e.g.,][]{dressler80, haynes86, butcher78}, and ram pressure stripping \citep{gunn72} is an important driver of galaxy evolution in clusters \citep[e.g.,][]{abadi99, Poggianti09, Cen14}. In the basic model, it acts by removing gas from the outside of the galaxy inwards (outside-in stripping), while leaving the existing stellar disk intact---as the galaxy moves into areas of denser intracluster medium (ICM), ram pressure increases and is able to strip the more tightly bound gas near the galaxy center. Modest starbursts may be triggered at the edge of the remaining gas disk \citep[e.g.,][]{koopmann04a, Crowl06}, and extraplanar star formation can occur in the tail of stripped gas \citep[e.g,][]{Cortese04, Sun07, hester10, Smith10, sun10,  Yagi10,  Abramson11, kenney14}. 

There is evidence that ram pressure stripping is an important star formation quenching mechanism at both low and high redshifts. Most of the nearest and clearest examples are found in the Virgo and Coma clusters. Ram pressure stripping is capable of almost completely stripping the interstellar medium (ISM) from spirals in the Coma cluster \citep[e.g.,][]{Smith10, Yagi10}, some of which have long tails of stripped ISM. Ram pressure can at least partially strip spirals in the less-massive Virgo cluster, can form long HI tails extending from an undisturbed old stellar disk \citep[e.g.,][]{Chung07}, and may completely strip dwarf galaxies \citep[e.g.,][]{kenney14}. Ram pressure stripping also shapes galaxy evolution at higher redshifts---for example, the recent spectroscopic study of a cluster at z = 0.31 by \citet{Rodriguez14} finds a population of disky k+a galaxies being quenched by an ICM-ISM interaction, either by ram pressure stripping material directly from the disk or by starvation removing the halo gas. \citet{ebeling14} describe six examples of ``jellyfish" galaxies with long, one-sided debris trails and evidence of triggered star formation at z$>$0.3. \citet{yagi15} find six galaxies at a cluster at z = 0.4 with extended H$\alpha$ tails resembling those of ram pressure stripped galaxies in the Coma cluster. \citet{Rauch14} detect galaxies in a Lyman-$\alpha$ halo at z = 3.2 with head-tail structures suggestive of ram pressure stripping. 

These recent studies provide increasing evidence that ram pressure stripping is a relatively common phenomenon in high-density areas throughout the universe, but its efficiency and the physical processes that accompany it are still not well-understood, partially due to a lack of high-resolution observations of stripping in action. In particular, we want to understand its importance as a star formation quenching mechanism, both in dense environments by directly stripping ISM from disks, and in less-dense environments by stripping gas from the halo, which results in starvation. Much is still unknown about the physics of ram pressure stripping and its effects on the complex, multi-phase and multi-density ISM. The basic \citet{gunn72} formula involves pressure due to the relative motion of the ICM and ISM pushing parcels of ISM gas out of a galaxy's gravitational potential well. However, momentum transfer is far from the only physical process at work \citep[e.g., ][]{weinberg14}. Simulations that include hydrodynamical effects produce the most realistic-looking results \citep[e.g.,][]{Roediger06}, but the resolution is not high enough to predict the features we see in HST images. Heating and cooling of the ISM play a major role in how gas is stripped \citep{tonnesen10,weinberg14}, and the presence of magnetic fields can affect the morphology of the stripped tail \citep[e.g.,][]{Ruszkowski14, Tonnesen14} and possibly bind together denser and less-dense gas, making the latter harder to strip \citep{kenney15}.

Understanding the effects of ram pressure stripping on very dense, hard-to-strip clouds is of special importance, since the densest clouds are where stars form. Dense clouds can decouple from the surrounding less-dense ISM \citep{Abramson14}, remaining in the disk as the less-dense material is swept away by the ICM wind, and could potentially start forming stars in the future. The process of dense cloud decoupling is directly related to the efficiency of ram pressure stripping, and determining the efficiency of ram pressure stripping is critical if we are to understand how star formation in galaxies is quenched as they fall into clusters, groups, and haloes. The quenching timescale of satellite galaxies falling into a cluster is a subject of much recent inquiry \citep[e.g.,][]{Wetzel13, Cen14, Muzzin14, haines2013}. More information on how efficiently star formation is quenched in infalling galaxies would provide needed constraints on the process, and high-resolution observations of nearby galaxies provide valuable information about which physical processes are taking place. 

In this work, we use HST images of stars and dust extinction in the Virgo Cluster galaxies NGC 4522 and NGC 4402 to constrain the effects of ram pressure stripping on both small and large spatial scales. These galaxies are among the best cases of active ram pressure stripping in the nearby Virgo cluster, which is the nearest cluster where clear examples of ram pressure stripping are observed, affording the opportunity to study this process in the highest-resolution study ($\sim$10 pc) to date---there are almost no studies of how ram pressure stripping affects the ISM on spatial scales of several tens of pc, with the exception of \citet[hereafter Paper 1]{Abramson14}. This work is also one of only a few studies of the effects of ram pressure stripping on ISM in the disk as well as in stripped tails. In Paper 1, which also examined NGC 4522 and NGC 4402, we describe the range of sizes and morphologies of dust clouds that have decoupled from the surrounding less-dense ISM. It is obvious from the images in that paper that the ISM has significant substructure related to ram pressure stripping on spatial scales from ten pc to several kpc. It is instructive to compare the two galaxies because we are observing different parts of them with respect to the ICM-ISM interaction. Extincting dust clouds are only visible if they are on the near side of the galaxy, between us and most of the stars in the disk. NGC 4522 is moving away from us in the cluster, so only the trailing side of the interaction is visible in dust extinction, and NGC 4402 is moving toward us in the cluster, so only the leading side is visible.

In the current work, we further examine the actual physical processes at work during stripping by considering the dust structures visible in the HST images in the context of observations at other wavelengths (HI, H$\alpha$, and total and polarized radio continuum), simulations, stellar population analyses, and information about the galaxies' ICM wind angles and rotation that already exists in the literature. We catalog and explore interesting kpc-scale features in the galaxies, including dust and stellar upturns curving out of the disk roughly in the direction of the ICM wind, extraplanar star-forming regions, and long, smooth dust fronts. By looking at the small-scale dust substructure and comparing the structures visible in the HST maps to those visible at other wavelengths, we can learn about the effects of ram pressure stripping on the complex, multi-phase, multi-density ISM. Observations and data reduction of the HST data, as well as a summary of previous work by other groups on these galaxies, are described in Paper 1. Section \ref{summaryofwork} includes a brief summary of the two galaxies, including results from Paper 1. In Section \ref{obsreduction}, we describe the observations and data reduction of the non-HST data used in this paper. In Section \ref{pontification}, we analyze dust features and emission at other wavelengths to constrain the physical processes that created the structures. Section \ref{simdiff} contains a discussion of the similarities and differences between the two galaxies and a comparison with NGC 4921, a stripped Coma spiral. In Section \ref{conclusions}, we present our conclusions.

\section{Summary of Existing Work}
\label{summaryofwork}

The Virgo spirals NGC 4522 and NGC 4402 are both excellent examples of active ram pressure stripping, and by comparing them we can develop a fuller picture of how the ICM-ISM interaction affects the ISM on small scales. Apart from the evolutionary stage and viewing angle of the ICM-ISM interaction, NGC 4522 and NGC 4402 are similar in many respects. They have similar optical luminosities ($M_B=$ -18.11 and -18.55, respectively) and high inclinations (78$\degree$ and 80$\degree$, respectively). They have similar HI masses and are both strongly HI-deficient for their types \citep{chung09}. Both have strong signatures of ram pressure stripping, including one-sided extraplanar HI and radio continuum tails \citep{kenney04, vollmer04, crowl05, vollmer10}, radially truncated star formation \citep{Koopmann04b, crowl05}, and polarized radio continuum ridges \citep{vollmer04, vollmer07, vollmer08} at the leading sides of the ram pressure interactions. 

The primary differences between the two galaxies are their stripping histories and their viewing angles. Star formation in the outer disk of NGC 4522 was quenched 50--100 Myr ago \citep{Crowl08}, while the outer disk of NGC 4402 was quenched 200--400 Myr ago. NGC 4522 has more extraplanar ISM (as traced by HI) and star formation (as traced by H$\alpha$) than NGC 4402 (40\% and 10\% of HI and H$\alpha$, respectively, versus 6\% and 0.3\%). NGC 4522 also has a smaller HI truncation radius (0.4 $R_{25}$ versus 0.6--0.7 $R_{25}$ in NGC 4402). The radio deficit parameter \citep{murphy09}, a proposed metric of ram pressure strength, is high in both galaxies, but it is twice as large in NGC 4522 as it is in NGC 4402 (0.096 vs. 0.045), indicating that NGC 4522 is experiencing stronger ongoing pressure than NGC 4402.  The above observables are all consistent with NGC 4522 experiencing stronger, more recent ram pressure than NGC 4402.  NGC 4402 is half as far from the cluster core (in projection; the three-dimensional distances may be different) as NGC 4522 is (0.4 Mpc vs. 0.8 Mpc), and since the ICM is generally denser closer to the cluster core, we might expect NGC 4402 to be experiencing stronger pressure.  However, a local density enhancement in the ICM near NGC 4522 may be responsible for its recent strong stripping \citep{kenney04}.

Viewing geometry causes us to observe different parts of the two galaxies with respect to the ICM-ISM interaction, and in Paper 1, we discussed the viewing geometry of the ICM-ISM interaction in detail. Stripping will be strongest on the leading side of a galaxy, where the ISM is exposed most directly to the ICM. However, we can only detect obscuring ISM tracers, such as dust, when they are between us and most of the starlight from the disk. NGC 4522 is moving away from us in the cluster, so only the trailing side of the interaction is visible in dust extinction, and NGC 4402 is moving toward us in the cluster, so only the leading side is visible. The stripped material will tend to be pushed downstream of the disk midplane. Figures \ref{4522regions} and \ref{4402regions} show the two galaxies with the locations of the regions discussed in this paper, along with the projected ICM wind direction and the galaxy's direction of rotation. The location of the small, decoupled clouds in the galaxies' partially stripped transition zones is shown in Figures 11 and 12 of Paper 1. The viewing geometry is also shown as a diagram in Figure 19 of Paper 1. In Paper 1, we speculate that the clouds currently on the visible (trailing) side of NGC 4522 probably decoupled when they were on the leading side, $\sim$1/2 an orbital period ago, which helps explain why there are fewer decoupled transition zone clouds in NGC 4522 than NGC 4402---in NGC 4522, the ICM wind has had $\sim$200---300 Myr to destroy clouds or push them out of the disk plane, but in NGC 4402, we are able to see leading-side clouds that decoupled more recently. 

In Paper 1, we began the work of characterizing sub-kpc scale dust features in the galaxies by examining the properties of a population of isolated dust clouds with sizes similar to giant molecular clouds (GMCs), located just beyond the main dust truncation radii of both galaxies in a partially-stripped ~2 kpc annulus of the disk we term the ``transition zone." We calculated a lower limit on the HI+H$_2$ masses of the isolated clouds using their dust extinctions and standard dust-to-gas ratios, and we found that a correction factor of $\sim$10 gave cloud masses similar to those measured in CO in the Milky Way and nearby galaxies for clouds of similar diameters. The discrepancy is probably due to the complicating factors of foreground light, resolution limitations, and cloud substructure, although it is possible that some of the clouds are less dense than GMCs. Using the corrected masses, we estimate that only a small fraction of the pre-stripping HI+H$_2$ ISM mass ($\sim$1--10\%) remains in the transition zones. Based on H$\alpha$ measurements, we found that a similar fraction ($\sim$2--3\%) of the pre-stripping star formation persists in the transition zones.

\section{Observations and Data Reduction}
\label{obsreduction}
\subsection{HST Imaging}
HST observations of NGC 4522 and NGC 4402 were fully described in Paper 1, and we give only a brief summary here. Both galaxies were observed with the Advanced Camera for Surveys on board HST. Images were taken of both galaxies in the F435W, F606W, and F814W bands (hereafter B, V, and I). Total exposure times in the three filters for NGC 4522 were 7356 s, 2358 s, and 2268 s, respectively, and exposure times for NGC 4402 were 7678 s, 2337 s, and 2372 s. Standard pipeline processing and alignment of the images were performed, and the individual exposures in each band were drizzled together to produce the final image for each band. 

\subsection{Optical Narrowband Imaging}
The H$\alpha$ + [NII] images (hereafter H$\alpha$) observations of NGC 4522 used in this work were first described in \citet{kenney99}. Exposures were taken at the WIYN 3.5 m telescope at KPNO in April 1997 using the S2KB CCD with a plate scale of 0$\arcsec$.2. A 30-minute exposure was performed using a narrowband filter with a bandwidth of 70$\AA$ centered on 6625$\AA$. The seeing was 1$\arcsec$.1 and standard IRAF packages were used to process, continuum-subtract, and combine the images. 

The H$\alpha$ observations of NGC 4402 used in this work were first described in \citet{crowl05} and were taken at the WIYN 3.5 m telescope using the MiniMosaic images. The observations were taken in April 2000 and the seeing was approximately 1$\arcsec$. Standard IRAF packages were used to create the final images. 

\subsection{HI 21 cm Line and 20 cm Continuum Observations}

NGC 4402 was observed with the VLA in C configuration to obtain HI 21 cm line and 20 cm radio continuum observations described in \citet{crowl05} and \citet{chung09}. Standard processing steps were followed in AIPS, producing images with a beam size of 17$\arcsec$.4 $\times$ 15$\arcsec$.2.  

NGC 4522 was observed with the VLA in CS configuration to obtain HI 21 cm line observations described in \citet{kenney04} and \citet{chung09}. Standard processing steps were followed in AIPS, producing an image with a beam size of $22\arcsec.5 \times 16\arcsec.4$. Radio continuum observations at 6 cm were originally presented in \citet{vollmer04} using VLA D configuration to produce an image with $15\arcsec \times 15\arcsec$. 

\begin{figure}[htb] 
 \centering
 \includegraphics[width=3.5in]{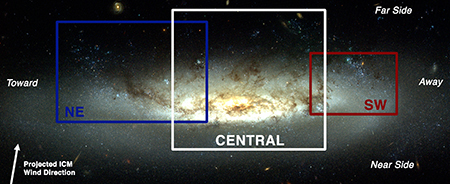} 
  \caption{Key to regions of NGC 4522, with near and far sides of the disk plane labeled. The galaxy is moving away from us in the cluster, so optical dust structures are on the trailing side of the ICM-ISM interaction. The sides rotating toward and away from us are also labeled.}
  \label{4522regions}
\end{figure}

\label{elongated}
\begin{figure}[htb] 
  \centering
  \includegraphics[width=3.5in]{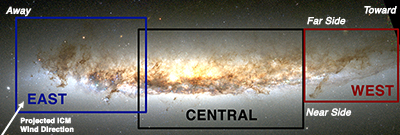} 
  \caption{Key to regions of NGC 4402, with near and far sides of the disk plane labeled. The galaxy is moving toward us in the cluster, so optical dust structures are on the leading side of the ICM-ISM interaction. The sides rotating toward and away from us are also labeled.}
  \label{4402regions}
\end{figure}

\section{Multi-wavelength Analysis of Key Structures}
\label{pontification}

\begin{figure*}[htbp] 
  \centering
  \includegraphics[width=6in]{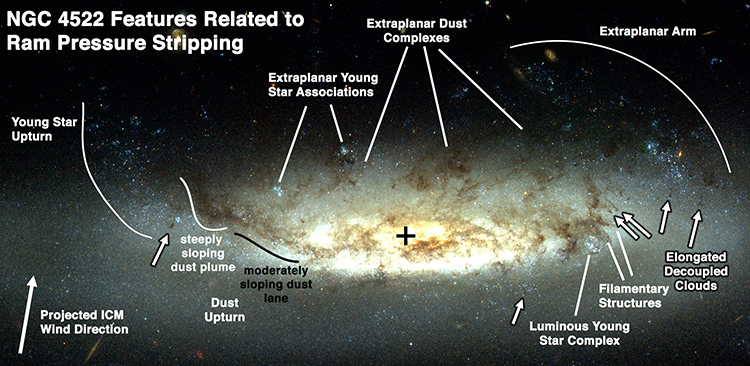} 
  \caption{Structures in NGC 4522 that are likely related to ram pressure stripping.}
  \label{4522rpsfeatures}
\end{figure*}

\begin{figure*}[htbp] 
  \centering
  \includegraphics[width=6in]{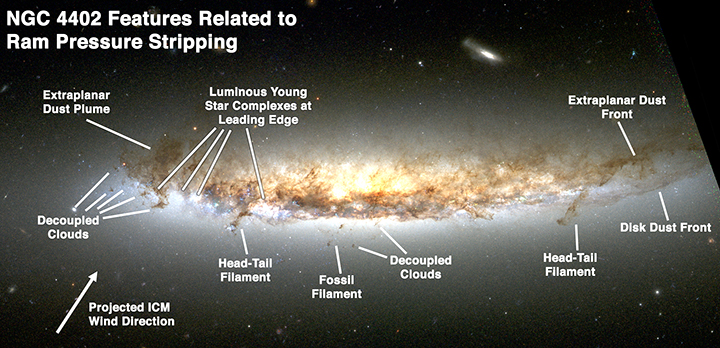} 
  \caption{Structures in NGC 4402 that are likely related to ram pressure stripping.}
  \label{4402rpsfeatures}
\end{figure*}

In this section, we discuss the multi-wavelength morphology and origins of key structures in NGC 4522 and NGC 4402. We analyze individual structures with the goals of understanding how a combination of ram pressure and pre-existing substructure caused them to form and classifying distinctive features in the two galaxies that have similar origins. Figures \ref{4522rpsfeatures} and \ref{4402rpsfeatures} show the structures whose characteristics and origins we discuss in this section. We compare some of these distinctive features to similar structures found in other stripped galaxies. The morphology of these features helps to constrain the physical processes acting on the disk during an ICM-ISM interaction and provides answers some key questions about the stripping process: Are detectable coherent chunks of the ISM ever pushed out at once? Does the ICM wind blow through holes in the disk ISM? Does ISM gas need to be shock-heated and ionized in order to be stripped (e.g., \citealt{weinberg14})? 

The formation and evolution of ram-pressure-stripped tails are a subject of intense interest for observers and theorists. The extraplanar material in NGC 4522 and NGC 4402 provides us with the highest-resolution view to date of stripped tails, allowing us to address several important questions raised by simulations and lower-resolution observations. For instance, under what conditions do dense substructures form in the stripped tail, and do they become dense enough to form stars? Can dense regions decouple from the surrounding less-dense material even after the gas has been removed from the disk plane? 

The evolution of the tail is also something we explore in this section. To what extent does the tail width reflect the outside-in progression of stripping in a galaxy? For instance, the Norma Cluster spiral ESO 137-001 \citep{sun2007, sun2010} has several ``orphan" HII regions downwind of it that have no accompanying X-ray or H$\alpha$ emission, and studying the structures in NGC 4522 and NGC 4402 might help us figure out how the orphan HII regions formed.  

For ease of description, we divide each galaxy into three large regions, the center and the two sides of the disk: NGC 4522 is divided into the northeast (NE), Central, and southwest (SW) (Figure \ref{4522regions}), and NGC 4402 is divided into the East, Central, and West (Figure \ref{4402regions}).

\subsection{NGC 4522 Upturn Morphology and Interpretation}
\label{upturnmorph}

\begin{figure*}[htb] 
  \centering
  \includegraphics[width=4in]{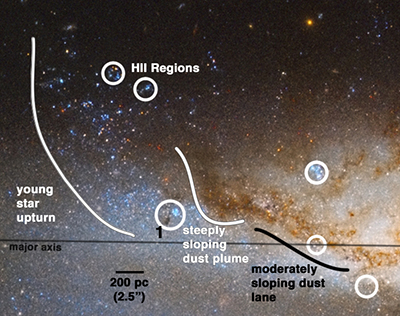} 
  \caption{NGC 4522 NE (upturn) region showing the two-component upturn structure, consisting of a dust upturn curving upward out of the disk and a stellar upturn located radially beyond the dust upturn. Star clusters corresponding to HII regions are circled. The stellar and dust upturns may represent different evolutionary phases of the stripping process. The two-component upturn structure is evidence that the active stripping zone has progressed radially inward over time. Also labeled is elongated dust cloud 1, whose position angle is discussed in Section \ref{swdecclouds}. Color HST image courtesy R. Colombari.}
  \label{4522nereg}
\end{figure*}

\begin{figure}[htbp]
  \centering
  \includegraphics[width=3.5in]{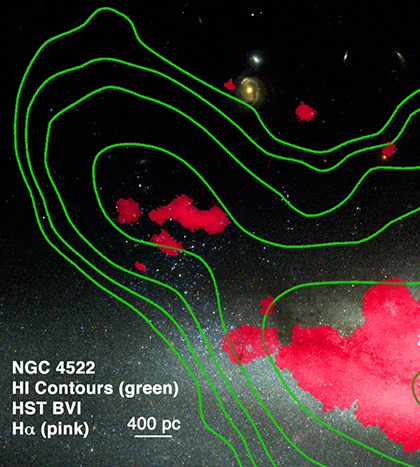} 
  \caption{NGC 4522 stellar and dust upturns in HST BVI, overlaid with HI contours (green) and H$\alpha$ distribution (pink). HI is coincident with both the stellar and dust upturns, and H$\alpha$ is preferentially associated with the dust upturn, indicating a lack of ongoing star formation in the stellar upturn except for the complex of HII regions at the top.}
  \label{4522-upturn}
\end{figure}

In NGC 4522, we have the opportunity to discover how the tail of stripped material is related to structures in the disk and how some of the tail's characteristics may have evolved from pre-existing disk substructure. The NE region of NGC 4522 (Figure \ref{4522nereg}) contains a large, continuous dust lane that curves out of the disk and extends into the extraplanar region (the ``dust upturn"), as well as a curved lane of young stars located radially beyond it (the ``young star upturn"), allowing us to see the connection between structures in the stripped tail and structures in the disk. The dust upturn is composed of the ``steeply sloping dust plume" and the ``moderately sloping dust lane." We refer to the whole curving structure of dust and stars as the ``upturn" and think that its origins are related to the ram pressure stripping process as the ISM is gradually pushed out of the disk plane from the outside in. In this section, we discuss the multi-wavelength morphology and possible origins of the striking upturn structure.

\subsubsection{The Dust Upturn} 
The dust upturn is a long, continuous-looking, two-part structure labelled ``steeply sloping dust plume" and ``moderately sloping dust lane" in Figure \ref{4522nereg}. The moderately and steeply sloping portions of the upturn have position angles (with respect to North in the images) of $358\pm10\degree$ and $324\pm10\degree$ respectively. The whole dust upturn structure begins 1.5 kpc (19$\arcsec$) from the galaxy center, extends to a 3.6 kpc (46$\arcsec$) radius along the major axis, and rises to a projected distance of at least 1.4 kpc (17$\arcsec$) above the major axis (downwind; see discussion in Section \ref{summaryofwork}). It is difficult to determine how far the dust upturn extends above the disk because there is insufficient background disk light to detect dust extinction beyond the dust upturn's maximum measured height. The galaxy's HI distribution extends an additional $\sim$1 kpc above the top of the dust upturn, indicating some extraplanar ISM in this region. The dust upturn cannot be simply a spiral arm within the disk plane: both parts are likely extraplanar, since the steeply sloping dust plume and a significant portion of the moderately sloping dust lane are strongly extincting and projected against the far side of the disk, above the major axis (shown as a horizontal line in Figure \ref{4522nereg}).

\subsubsection{The Stellar Upturn}
Centered $\sim$900 pc further out radially from the steeply sloping dust plume, the stellar upturn is a wide lane of stars, including many young blue stars (labeled ``young star upturn" in Figure \ref{4522nereg}), that leaves the disk at the same angle that the steeply sloping dust plume does ($325\degree\pm10\degree$ for the stellar upturn, and $324\degree\pm10\degree$ for the steeply sloping dust plume). The stars that are highest above the major axis are clearly extraplanar. The stellar upturn extends to a height of 2.8 kpc (35$\arcsec$) above the major axis.

The top of the stellar upturn has both HI and HII emission, showing the presence of extraplanar ISM and star formation (Figure \ref{4522-upturn}). There are three distinct HII regions along the top of the stellar upturn 2.2 kpc above the major axis, as well as some fainter, diffuse H$\alpha$ emission just below them. Between the base and the top of the stellar upturn, only a small amount of diffuse H$\alpha$ is detected. Although there are no HII regions in most of the upturn base, there is one significant HII region at an intermediate radius between the stellar upturn and the dust upturn, just above the major axis and near a decoupled transition zone dust cloud. The relative locations of the HI and H$\alpha$ throughout the upturn region (Figure \ref{4522-upturn}) suggest that the extraplanar ISM that originally formed the stellar upturn has now been largely stripped away, except for a blob of ISM that remains at the very top. Although some of the stellar upturn is inside the outer HI contours, the HI beam size is very large, so the upturn being inside the outermost HI contours does not necessarily indicate the presence of HI. 

\subsubsection{Upturn Interpretation}
The morphology of the dust upturn (Figure \ref{4522nereg}) appears to show the effects of the ICM wind as the ISM is pushed out of the disk from the outside in, and this morphology strongly suggests that large ($>$1 kpc) chunks of the cold ISM can be pushed out together---it need not all be ionized before being pushed out, as suggested by \citet{weinberg14}. Assuming that the dust upturn is in fact a continuous structure, ram pressure seems to be affecting the two parts differently. The continuity of the moderately sloping dust lane and the lack of dust clouds below it seem to indicate that the ICM has not broken through the ISM here, but is pushing it all out of the disk together. Meanwhile, the steeply sloping dust plume is at the edge of the stripped portion of the disk where the ICM is presumably flowing freely, so it may be exposed to a higher-velocity flow that pushes it out of the disk at a steeper angle. 

The stellar upturn is located radially outward from the dust upturn, and we propose that the active stripping zone has progressed radially inward over time and used to be where the stellar upturn is now located. It is likely that the ISM that used to surround the young stars in most of the upturn has now been stripped and pushed further upwards, leaving behind a trail of young stars, and the only gas left now is at the top of the stellar upturn. Rather than coming from the gas that used to be in the stellar upturn region, the dust upturn formed from material that was stripped from the remaining gas disk more recently than the ISM that created the stellar upturn. In contrast to the many bright young stars of the stellar upturn, there appears to be little ongoing star formation in the dust upturn, based on 8-$\mu$m and 24-$\mu$m Spitzer observations (Kenney et al., in prep). Since the material that formed the stellar upturn appears to have resulted in more star formation post-stripping than the dust upturn likely will, it is possible that the stellar upturn is a result of the ram pressure stripping of a denser region, such as a spiral arm. In this scenario, we are now witnessing the stripping of a less-dense region, such as an inter-arm region, whose material is becoming the dust upturn. If so, this observation shows a connection between substructure in the ram-pressure-stripped tail and pre-existing substructure in the disk plane.

 \begin{figure}[htbp] 
  \centering
  \includegraphics[width=3.5in]{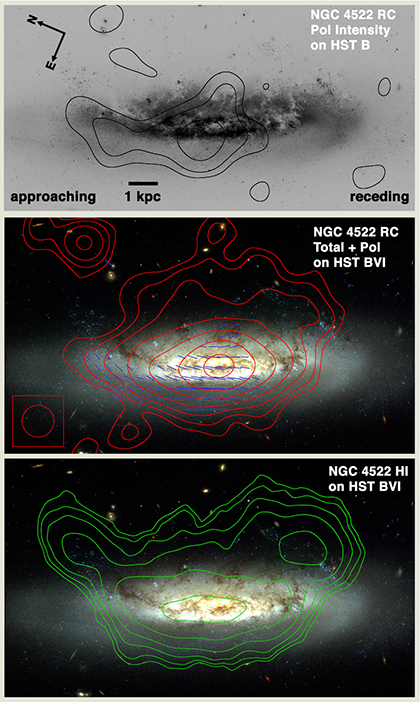} 
  \caption{NCC 4522 HST B image with polarized RC intensity contours at 1E-5 x 2, 4, 8... Jy/Beam; NGC 4522 color optical image with RC intensity (red contours) at 1, 2, 4, 8...$\times$1E-4 Jy/Beam and RC polarization (blue lines) at 1 arcsec = 1E-5 Jy/beam; HI (green contours) at 0.4 x 2, 4, 8... mJy. Approaching and receding sides of the major axis are labelled. 
} 
   \label{4522rc}
\end{figure}

\begin{figure}[htbp] 
  \centering
  \includegraphics[width=3.25in]{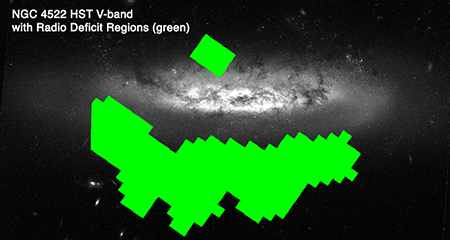} 
  \caption{NGC 4522 radio deficit regions (green) on HST V-band.}
  \label{4522-radiodef}
\end{figure}

\subsubsection{Upturn Radio Continuum Morphology \& Interpretation}

The location and morphology of the radio deficit region in NGC 4522 \citep{murphy09} are consistent with the interpretation that all of the ISM in the moderately sloping dust lane is being pushed up together. Total radio continuum emission consists of both polarized and unpolarized emission, the latter resulting from cosmic ray electrons gyrating around ISM magnetic fields that are disordered on small scales. Ram pressure sweeps magnetic field lines and cosmic ray electrons away from the outer halo, creating a local radio deficit region in the halo where the total radio continuum emission is weaker than would be expected based on the observed FIR distribution and the FIR-radio correlation in other galaxies \citep[e.g.,][]{murphy09}. In addition to the radio deficit region in the halo, NGC 4522 also includes linearly polarized radio continuum ridges, indicating a magnetic field direction parallel to the ridge (Figure \ref{4522rc}; \citealt{vollmer04}). 

In order for radio continuum emission to be polarized, a large-scale, ordered magnetic field must be present in either the disk or the halo, and if the ICM were flowing freely through the disk plane, a large-scale magnetic field would likely be disrupted. Therefore, the polarized radio continuum emission is consistent with the above interpretation that the ICM has yet to break through the ISM in the moderately sloping upturn. It is possible that the upturn region contains no clouds dense enough to decouple and resist stripping. 

The morphology of the polarized radio continuum emission near the dust upturn supports the idea that ICM pressure is causing the ISM to pile up in a ridge-like structure at the interaction boundary. A ridge of polarized radio continuum emission is centered $\sim$900 pc below the disk from the center of the dust upturn (Figure \ref{4522rc}) and may extend all the way to the dust upturn. The ridge of polarized radio continuum emission curves along the outside of the dust distribution from the upturn region in the NE, extending to near the galaxy center. This prominent ridge of polarized radio continuum emission below the upturn suggest a buildup of magnetic fields and relativistic electrons along the bottom of the upturn, where the ICM wind has not yet pushed most of the ISM out of the disk.

Just upwind (NE) of the polarized radio continuum emission, the deficit region curves upward toward the disk plane with roughly the same angle as the dust upturn (Figure \ref{4522-radiodef}; also \citealt{murphy09}, Figure 5). The radio deficit region appears to be in the halo and perhaps also the outer disk of the galaxy, along the outer edge of the detected HI distribution. The location and curvature of the radio deficit region is further evidence that the dust upturn is near the leading side of the ICM-ISM interaction: the dust upturn is closest to the remaining ISM disk, the stellar upturn is along its outer edge, the ridge of polarized radio continuum is centered 900 pc below the dust upturn, and the radio deficit region is located beyond the polarized ridge.

\subsection{NGC 4522 Upturn Comparison With Other Galaxies}
\label{upturncomp}

In this section, we compare the stellar and dust upturn in NGC 4522 (Figure \ref{4522nereg}) to distinctive structures in other galaxies that are also thought to result from the ram pressure stripping process. Such structures, in which the ISM shows evidence of being progressively stripped from the galaxy, are present in several spiral and dwarf galaxies in the Virgo and Norma clusters \citep[e.g.,][]{Sun07, sun10, Jachym14, kenney14}. To constrain the physical processes that occur during an ICM-ISM ram pressure interaction, it is valuable to compare the structures that exist in these galaxies. 

The stellar upturn with gas and star formation at its head is morphologically similar to the ``fireball" structures observed trailing the Virgo cluster dwarf galaxy IC 3418 \citep{kenney14} and likely has a similar origin. The stripped gas has progressed downwind of the galaxy, leaving behind a trail of newly-formed stars that decouple from the accelerated ISM. A linear structure of young stars is created as the gas cloud continues to be accelerated by ram pressure, leaving no gas between the newly-formed stars and the galaxy. This type of structure, with an outer gas head and a tail of stars extending toward the galaxy, is seen on both large and small spatial scales in the tail of IC 3418. In the IC 3418 fireball structures, the newly-formed stars are visible in the UV and optical, and there is a small amount of HI coincident with several HII regions at the outer end of the tail, nearly 17 kpc from the main body of the galaxy. At ground-based resolution similar to the optical images of IC 3418, NGC 4522's stellar upturn with HI and HII regions at its top would resemble a shorter version of the the tail of IC 3418. The narrower optical trails in IC 3418 have a characteristic length and width of 0.5--3 kpc and 200--400 pc, respectively. The NGC 4522 stellar upturn extends a comparable 2.8 kpc above the major axis, and its width of $\sim$1 kpc is wider than the optical structures in IC 3418.  

There are some differences in morphology between the tail of IC 3418 and the ISM/stellar upturn of NGC 4522. A major difference is the stellar features' orientations with respect to the ICM wind direction. The overall orientation of the large-scale tail in IC 3418 matches that of the small head-tail linear stellar features, which is presumably the global wind direction. In NGC 4522, where the global ICM wind direction can be constrained by the position angles of the elongated transition zone clouds \citep{Abramson14} and the location of the radio deficit region, the angle of the stellar upturn (325$\degree\pm10\degree$) is offset from that of the global wind direction (288$\degree \pm 10\degree$) by 37$\degree\pm20\degree$. Instead, the angle of the stellar upturn may reflect the local ICM flow direction at the time when the stellar upturn formed. The flow of the ICM around the remaining gas in the disk is complicated \citep[e.g.,][]{Roediger06, tonnesen10}, and the local ICM flow direction near the disk will not always be in the projected large-scale wind direction. Since the dust upturn leaves the disk at a similar angle to the stellar upturn and probably formed more recently, the local flow in this region has probably been in a similar direction for a significant amount of time (at least $\sim$10--20 Myr if the stars in the stellar upturn formed recently). 

Another difference between the tail in IC 3418 and the ISM/stellar upturn in NGC 4522 is that the upturn lacks stellar substructure, while the tail has trails of young stars extending downstream. There are no obvious stellar streams trailing the upturn HII regions. The lack of substructure in the stellar upturn argues that the upturn may be more analogous to an individual fireball than to the entire tail of IC 3418.
 
There are good reasons that the stripping process may not be identical for the two galaxies, since the dwarf galaxy IC 3418 is less massive and in projection is located in a denser part of the cluster, near M87 (277 kpc; \citealt{McLaughlin99}), where ram pressure would be stronger. The main body of IC 3418 appears to be completely stripped, while there is still significant gas remaining in the disk of NGC 4522. Although IC 3418 is clearly at a later evolutionary stage of stripping than NGC 4522, the two galaxies' tails probably formed in similar ways. 

There is similar evidence for outside-in stripping in the Norma Cluster spiral ESO 137-001 \citep{Sun07, sun10}. The galaxy has two narrow tails with X-ray and H$\alpha$ emission, with CO emission also present in the secondary tail, as well as a number of orphan extraplanar HII regions that form a broad tail but are not associated with X-ray or diffuse H$\alpha$ emission. The origin of the broad tail is not well-understood, but we think they may be similar to structures in NGC 4522 and IC 3418. \citet{Jachym14} propose that stripping progresses from the outside of the galaxy inwards, forming a broader tail at early stages, which narrows as the radial extent of the remaining gas disk shrinks. Thus, we see the orphan HII regions where there was once a broad ISM tail made up of gas stripped from the outer disk at an earlier stage of stripping. The HII regions' spatial distribution shows where the densest ISM was found in an earlier evolutionary stage of stripping. Now, the orphaned HII regions have decoupled from the less-dense surrounding ISM, which has been pushed further downstream. With no more gas in the outer disk to resupply the broad tail, there are now orphaned HII regions in a broad tail outside of the current narrow tail of stripped ISM.

We believe that the stellar upturn forms a broad tail of extraplanar stars in NGC 4522. The dust upturn, located radially inside the stellar upturn, is an indication of the current extent of the narrow tail. Over time, the upturn structure in NGC 4522 may evolve into something with a similar structure to the broad tail in ESO 137-001 if the HII regions at the top of the upturn decouple from the surrounding less-dense ISM.

An upturn structure similar in some ways to the one in NGC 4522 has been identified in the Virgo Cluster spiral NGC 4330 \citep{Abramson11}, visible in dust and young stars curving downwind away from the major axis at a fairly steep angle. The most prominent feature of this upturn is a large, continuous, highly extincting dust plume extending 1.5 kpc downwind from the major axis. Some evidence was found for a concentration of blue stars along the outer edge of the dust plume, and it is possible that the galaxy contains a stellar upturn located radially outside of the upturn dust plume, although the ground-based data was inconclusive. Like the upturn in NGC 4522, the upturn in NGC 4330 also seems to be the result of the ICM wind pushing ISM out of the disk at gradually smaller radii. There are some key differences, such as the fact that the NGC 4330 upturn does not seem to have shallow and steep components or a separate stellar upturn.  

In summary, in NGC 4522, IC 3418, ESO 137-001, and NGC 4330 all show an evolution of the ISM distribution as stripping progresses---broad tails tend to get narrower, and material is pushed progressively further from the disk plane over time.

\subsection{NGC 4522 Central and Southwest Regions: The Extraplanar Arm}
\label{xparm}

\begin{figure*}[htb] 
  \centering
  \includegraphics[width=5in]{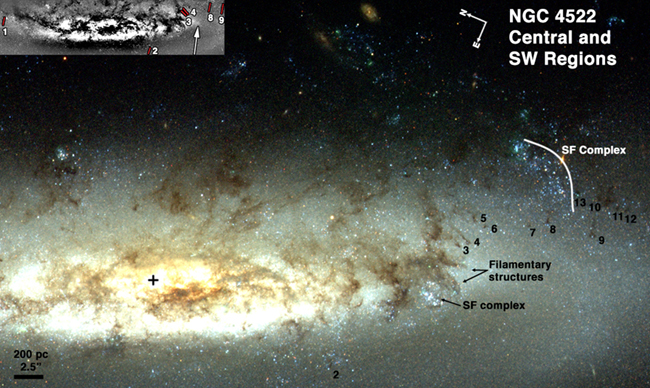} 
  \caption{NGC 4522 central and SW regions showing decoupled dust clouds 2--13 and the extraplanar stellar arm structure. The inset shows the position angles of the galaxy's elongated clouds---note that outermost clouds 1, 2, 8, and 9 have the same position angle, which we interpret as the galaxy's global ICM wind angle.}
  \label{4522swreg}
\end{figure*}

The SW and central regions of NGC 4522 (Figure \ref{4522swreg}) have a very different dust and stellar morphology from that of the NE (Figure \ref{4522nereg}), showing that other kinds of structures may also form from stripped material. The SW side has been affected differently by the stripping process, probably because of the interaction between the galaxy's rotation and ram pressure, which are discussed later in this section. The SW side of the galaxy includes a mostly-stripped disk plane beyond a 3.4 kpc radius with dust clouds at varying position angles. Above the stripped disk plane is an extraplanar arm of stripped material and star formation, also detected in HI and CO \citep{vollmer08}, that appears to be a distinct kinematic and morphological feature. We define the extraplanar arm as the region of dust and stars above the central and SW regions of the disk shown in Figure \ref{4522sw-wind}, extending $\sim$5 kpc from near the center of the galaxy to the blue stars just beyond the SW decoupled clouds. In this section, we discuss the multi-wavelength morphology and possible origins of the structures on the SW side of the galaxy. 

\subsubsection{SW Decoupled Dust Clouds and ICM Flow}
\label{swdecclouds}
The SW region features a number of decoupled extraplanar dust clouds (Figure \ref{4522swreg}), near the bottom of the region we have called the extraplanar arm (further discussed below). These clouds were first described in Paper 1, where we measured their masses and noted that the elongated dust clouds (clouds 3, 4, 8, and 9) vary by 60$\degree$ in their position angles over a radial range of $\sim$1.5 kpc along the major axis. They are clearly extraplanar, because they are projected against the far side of the galaxy $>$1 kpc above the major axis but are highly extincting, indicating that they are in front of most of the star light (see Section \ref{summaryofwork} for further discussion of the wind angle and viewing angle). 

The outermost elongated clouds, Clouds 8 and 9 [$\sim$930 pc (12$\arcsec$) and $\sim$1.4 kpc (18$\arcsec$) SW of main dust lane, respectively] have position angles of $\sim$285$\degree$. This is similar to the position angle of two clouds elsewhere in the galaxy, Clouds 1 and 2 (in the NE and central regions, respectively), which are both located a distance of $\sim$400 pc (5$\arcsec$) outside of the main dust lane. Because this position angle is consistent in the outermost elongated clouds throughout the galaxy, we think it reflects the global projected ICM wind angle. 

Clouds 3 and 4 are located nearer to the galaxy center at projected distances of 3.4 kpc (43$\arcsec$) and 3.5 kpc (44$\arcsec$), just at the edge of the remaining ISM. Their position angle of $\sim$345$\degree$ is significantly different from the global wind angle. They are close to and possibly associated with the main dust distribution, suggesting that clouds very close to the main dust lane have position angles determined by something other than the global wind angle. They have dense head and elongated tail components, a morphology that suggests that they are being exposed to the ICM wind. 

We propose that an elongated cloud's position angle reflects the local ICM flow direction, and the local flow direction may differ from the global one near the edge of the remaining ISM disk due to hydrodynamical effects. We propose that Clouds 3 and 4 indicate the local ICM flow direction near the main dust distribution. They are at the edge of the main disk of ISM, as traced by dust and H$\alpha$, suggesting that they may be in the process of decoupling from the surrounding ISM. In this scenario, the movement of the ICM around the edges of the remaining gas disk creates complex local flows, while the flow characteristics further out is simpler and roughly in the global wind direction. The varying cloud position angles in NGC 4522 are evidence the ICM wind can act on the disk ISM from directions other than the global wind angle---it appears that clouds more than 400 pc from the dust lane have position angles that reflect the global wind angle.

\subsubsection{Extraplanar Young Stars and H$\alpha$}

The extraplanar arm structure contains a number of groupings of bright, blue stars, which we interpret as young stars, with associated HI and H$\alpha$ emission. There is a lane of young stars at projected heights ranging from 1.4 kpc (18$\arcsec$) above the major axis (at the top of the curve labeled ``SF complex" in Figure \ref{4522swreg}), to 0.7 kpc where the star lane ends above the stripped disk plane in the SW. Most of the denser groupings of extraplanar blue stars are located within the HI contours and are visible in H$\alpha$. Bright blue stars are coincident with the H$\alpha$ emission in the SW extraplanar region and are also interspersed between the H$\alpha$ peaks, suggesting a continuous structure of HII regions and young stars (HI and H$\alpha$ shown in Figure \ref{4522swhiha}, and the BVI structures are labeled in Figure \ref{4522sw-wind}). \citet{vollmer08} use an N-body sticky particle simulation of ram pressure stripping to attempt to match the HI morphology and kinematics of NGC 4522, and the models support the idea that the long grouping of stars could be part of an extraplanar arm of stripped material traced by CO and HI emission in the same region. 
 
The outermost extraplanar HII region in the SW (labeled in Figure \ref{4522swhiha}) is at a $\sim$1.5 kpc larger galactocentric radius than the decoupled dust clouds and is located 2.8 kpc above the major axis in projection. It is outside the galaxy's HI distribution, suggesting that most or all of the surrounding ISM has been stripped away. This HII region may be analogous to the broad tail of orphan HII regions that have decoupled from the lower-density ISM in the tail of Norma Cluster spiral ESO 137-001 (discussed in Section \ref{upturncomp}). We propose that in both ESO 137-001 and NGC 4522, ram pressure strips the outer disk first, leaving behind decoupled extraplanar HII regions outside the center of the stripped gas tail. In this scenario, the outermost HII region originated from an earlier evolutionary stage of stripping, from gas stripped further out in the disk than the current SW extraplanar ISM. 

\subsubsection{Extraplanar Clouds Between the Disk and Extraplanar Arm}

The extraplanar dust clouds are located in between the extraplanar arm structure and the disk. Did the extraplanar clouds form in situ from stripped material, or were they directly stripped at their current GMC-like densities? The two scenarios involve a similar process, because if the clouds did collapse in situ, they formed from overdensities that probably already existed when the gas was stripped. Whether these clouds originated in the disk plane or formed in situ above the disk plane, at some point they decoupled from the surrounding less-dense gas and were accelerated more slowly than the surrounding less-dense gas, which is why the clouds appear in an otherwise stripped part of the galaxy.

In general, ram pressure in Virgo is not strong enough to directly strip GMCs, but there is some uncertainty in our measurement of the cloud masses \citep[see][]{Abramson14} so these clouds may be less dense than typical GMCs. Regardless of whether they collapsed in situ or were directly stripped, these clouds (or the overdensities that collapsed to form them) represent the densest gas able to be directly stripped from the disk plane. If that is true, where did the gas dense enough to form stars in the extraplanar arm (which is further above the disk plane than the clouds) come from? If the SW extraplanar dust clouds are remnants of the most overdense regions that can be stripped from the disk plane, the clouds that collapsed to form the stars in the extraplanar arm must have collapsed from the less-dense stripped ISM. 

\subsubsection{Extraplanar Arm HI Kinematics}

The extraplanar arm is distinct kinematically and morphologically. The HI kinematics of the region containing the extraplanar arm are described by \citet{kenney04}, who note that the HI velocities in the SW extraplanar region are shifted by up to 100 km s$^{-1}$ toward the Virgo Cluster velocity compared to the disk HI. The velocity shift is consistent with acceleration of stripped gas by the ICM wind. The extraplanar HI velocities deviate most strongly from the disk velocities in the area coincident with the SW extraplanar arm (Figure \ref{4522hivel}). In contrast, the extraplanar ISM on the NE side has velocities within 10--20 km s$^{-1}$ of the disk-plane velocities. Above the extraplanar arm, the HI returns to velocities similar to those near the disk. This suggests that the extraplanar HI has two components along the line of sight: the extraplanar arm, and a more extended region of lower-density gas. The CO observations and models of this galaxy by \citet{vollmer08} support the existence of an extraplanar arm dense enough to support star formation extending toward us in the SW, along with a more extended HI distribution in the same region. 

The velocity shift of the extraplanar arm relative to the disk ISM may be related to the galaxy's rotation. The relative strength of ram pressure in a given part of the galaxy results from a combination of the global ICM wind velocity and the rotational velocity of a parcel of gas with respect to the ICM wind. \citet{kenney04} suggest that the differences in the velocity of the extraplanar material on the two sides arise from the fact that the SW is rotating into the ICM wind, while the NE is rotating away from it---the ram pressure in the SW may be up to a factor of 1.3 greater than that experienced by the NE. The enhanced ram pressure in the SW could also account for the larger amount of extraplanar ISM and associated star formation. 

In the simulation of \citet{vollmer08}, the upper end of the extraplanar arm has been most accelerated in the direction of the ICM wind, which is toward us along the line of sight. This is consistent with the observed blueshift of the extraplanar HI. The part of the extraplanar arm furthest from the disk plane should also have originated from material at the largest galactocentric radius, since ram pressure removes the ISM from larger radii first.

\subsubsection{Radio Continuum in the Extraplanar Arm}

The total radio continuum distribution on the SW side of the galaxy ends at roughly the same radius as the main body of the dust distribution (Figures \ref{4522rc}, \ref{4522-radiodef}). The total radio continuum emission has a distribution comparable to that of the HI, with some emission from the extraplanar arm region. The polarized radio continuum ends at an even smaller radius, within 1--2 kpc of the galaxy center, and we detect no polarized radio continuum emission from the extraplanar arm. There is no long, continuous ISM structure like the upturn in the NE---rather, there is a disk plane free of dust and numerous small extraplanar clouds, some of which are elongated, that appear to have partially or totally decoupled from the surrounding ISM. Some of them likely decoupled from the gas that now forms the extraplanar arm.  

We propose that in the SW, the lack of total and polarized radio continuum emission from the disk plane indicates that most of the ISM, including magnetic fields and relativistic electrons in the disk and adjacent halo, has been pushed into the region of the extraplanar arm. There are several possible reasons why we detect total but not polarized radio continuum emission from the extraplanar arm, including beam depolarization or a lack of large-scale ordered magnetic fields. The morphology of the SW isolated clouds suggests that the ICM wind is interacting directly with the individual dust clouds and flowing freely through the parts of the disk that have already been stripped. The clouds have been elongated by the ICM wind, either by hydrodynamical ablation or as a result of the wind stretching out the dense ISM along magnetic field lines. These possibilities are discussed more fully in Section \ref{magfields}.

\begin{figure*}[htb] 
  \centering
  \includegraphics[width=5in]{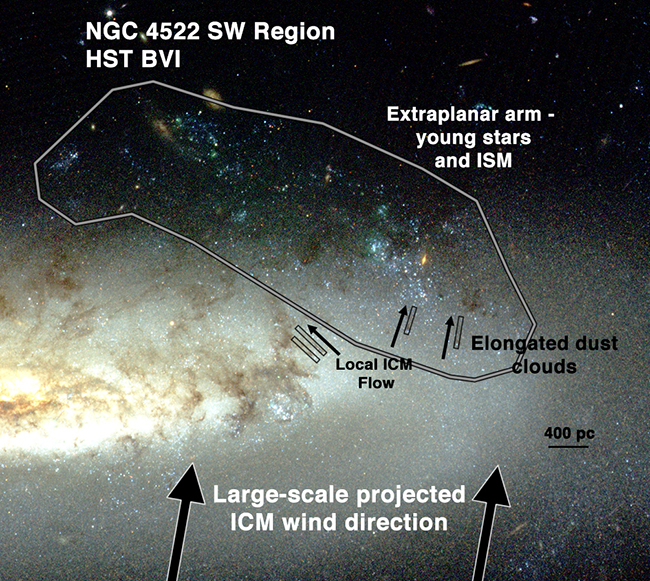} 
  \caption{NGC 4522 SW with HST optical, showing the extraplanar arm structure, position angles of the projected ICM wind, elongated dust clouds and decoupled dust clouds between the disk and extraplanar arm, and the possible local ICM flow through the SW disk. The local ICM flow direction may differ from the global one near the remaining ISM disk due to hydrodynamical effects.}
  \label{4522sw-wind}
\end{figure*}

\begin{figure}[htb] 
  \centering
  \includegraphics[width=3.25in]{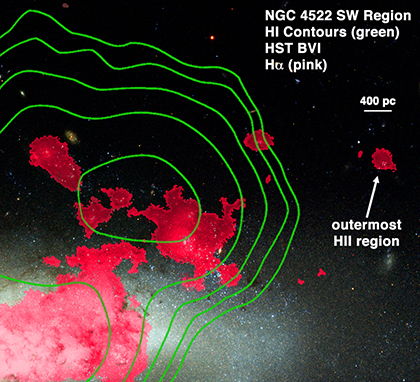} 
  \caption{Part of the NGC 4522 SW extraplanar arm region with HST optical, HI, and H$\alpha$.}
  \label{4522swhiha}
\end{figure}

\begin{figure}[htb] 
  \centering
  \includegraphics[width=3.25in]{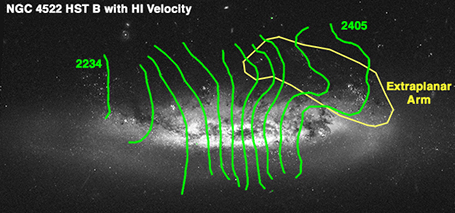} 
  \caption{NGC 4522 HST B with HI velocities (km/s) and location of extraplanar arm. HI velocity measurements indicate that the extraplanar arm is a kinematically distinct feature.}
  \label{4522hivel}
\end{figure}

\subsubsection{The Extraplanar Arm ISM Mass}
 
We can estimate the ISM mass of the extraplanar arm and compare it to the estimated pre-stripping mass in the part of the disk below it to discover what fraction of the ISM has remained near the galaxy after stripping. We are able to detect dust extinction to a height of $\sim$1 kpc above the major axis, although the dust clouds are probably only a small fraction of the pre-stripping mass. \citet{Abramson14} estimate a total mass of 5.6$\times10^{6}$ M$_{\odot}$ for the SW extraplanar dust clouds. The HI + H$_2$ mass in the arm is 1$\times10^8$M$_{\odot}$ based on the HI observations of Chung et al. (2009) used elsewhere in this paper and the CO observations of Vollmer et al. (2008). 

Paper 1 introduced the concept of the transition zone, the area of the disk in a stripped galaxy that has been stripped of all but the densest clouds, which in NGC 4522 extends from $\sim$3--4.5 kpc and encompasses a volume 1 kpc above and below the disk. We estimated a pre-stripping transition zone mass for the near side of the galaxy of 7.9$\times10^{7}$ M$_{\odot}$ by assuming an exponential HI + H$_2$ distribution based on the profile of the inner disk. The extraplanar arm is probably made up of material stripped from both the near and far sides of the galaxy and so potentially contains material from $\sim50\%$ of the transition zone. Assuming that the transition zone on the far side of the galaxy has a similar extent to that on the near side, the mass of the extraplanar arm matches the mass of the corresponding transition zone to within $\sim20\%$.

\subsubsection{Extraplanar Arm Structure Formation}

Several stripping simulations predict the formation of extraplanar arm structures in the increasing-pressure quadrant that is rotating into the ICM wind \citep{Schulz01, vollmer04b, Jachym09}, and examples have been observed in HI \citep{Phookun95, vollmer04b, vollmer08}. In previous examples of extraplanar arms, it was unclear whether the dense gas in the arm was stripped as dense clouds or was stripped in a less-dense form and later collapsed into dense clouds with the potential for star formation. The extraplanar arm structure in NGC 4522 appears different from other examples, because we view the galaxy nearly edge-on and can tell that the arm is clearly extraplanar, and HST resolution allows us to detect smaller dust clouds than had previously been observed in NGC 4522. Our observations reveal the key detail that the decoupled dust clouds lie just below the extraplanar arm, and so remain after the rest of the less-dense ISM was swept up into the arm region. It seems possible that the decoupled clouds represent the densest substructures in the pre-stripping ISM that were able to be stripped, since denser clouds will remain closer to the disk plane during the stripping process than less-dense gas; the rest of the extraplanar arm is gas that was stripped in a less-dense form and later collapsed. 
 
\subsection{NGC 4402: Elongated Filaments and Surrounding Regions}

We now describe and analyze the dust and stellar properties of NGC 4402, in which we view the stripping interaction from the leading side. Using a multi-wavelength data set similar to that of NGC 4522, we identify structures likely created by ram pressure stripping and discuss the physical processes that created them. We use previous works' constraints on the ICM wind angle and the galaxy's rotation.

\begin{figure}[htb] 
  \centering
  \includegraphics[width=3in]{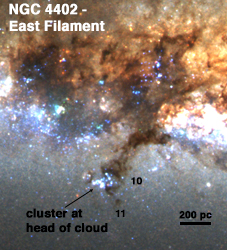} 
  \caption{Cloud 10, the NGC 4402 eastern dust filament, and nearby dust complexes and stellar associations.}
  \label{4402sefilament}
\end{figure}

\begin{figure*}[htbp] 
  \centering
  \includegraphics[width=5in]{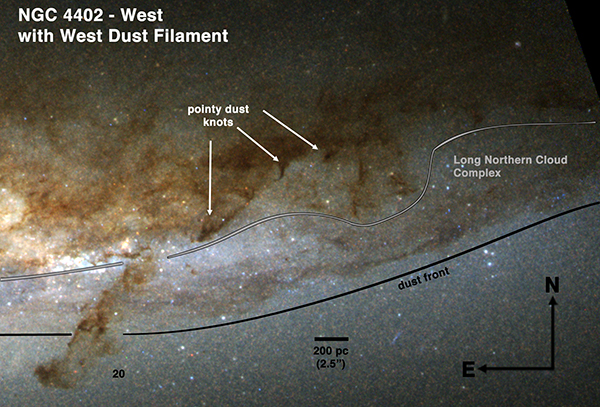} 
  \caption{NGC 4402 western region, including Cloud 20, the western dust filament.}
  \label{4402wfilament}
\end{figure*}

Perhaps the most prominent dust features in NGC 4402 are two elongated, kpc-scale dust filaments first described by \citet{crowl05}. Labeled here as Clouds 10 (Figure \ref{4402sefilament}) and 20 (Figure \ref{4402wfilament}), they are located in the transition zone just beyond the main dust disk near the disk plane at projected galactocentric radii of $\sim$4.5 kpc on the eastern and western sides of the minor axis. They are both much larger than the other discrete dust clouds in the transition zone, with lengths of 0.6 and 0.9 kpc respectively, and masses of $\sim10^7$ M$_{\odot}$ \citep{Abramson14}. Both clouds are have substructures that give clues to their origins and 3-dimensional morphologies. In this section, we discuss the filaments, their substructure, and their surroundings on the eastern and western sides of the galaxy.

\begin{figure}[htb] 
  \centering
  \includegraphics[width=3.5in]{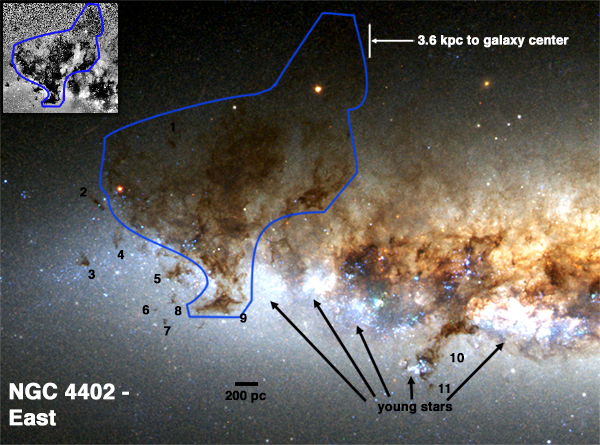} 
  \caption{NGC 4402 -- HST BVR image of the eastern section of the disk, highlighting the dust distribution and young stars. The inset is an unsharp masked image that more clearly shows the northernmost extinction in the dust plume.}
  \label{4402ereg}
\end{figure}

\begin{figure}[htbp] 
  \centering
  \includegraphics[width=3.25in]{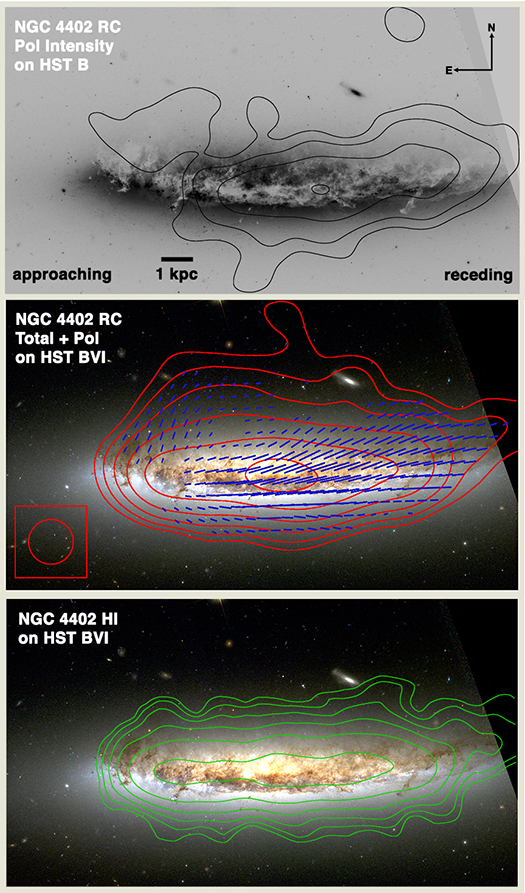} 
  \caption{NGC 4402 HST B image with polarized RC intensity contours at 1E-5 x 4, 8, 16... Jy/Beam; NGC 4402 color optical image with intensity (red contours) at 1,2,4,8... $\times$1E-4 Jy/Beam and RC polarization (blue lines) at 1 arcsec = 2.5 E-5 Jy/Beam; HI (green contours) at 0.4 x 2, 4, 8... mJy. Approaching and receding sides of the major axis are labelled.}
  \label{4402rc}
\end{figure}

\subsubsection{NGC 4402 Eastern Region---Active Star Formation, Large Elongated Filament}
\label{4402east}
The eastern region of NGC 4402 (Figure \ref{4402ereg}) appears to be interacting most directly with the ICM wind, and Cloud 10, numerous smaller decoupled clouds, and the large extraplanar dust complex outlined in blue in Figure \ref{4402ereg} are all evidence of disturbance by ram pressure. Cloud 10 has a luminous complex of young stars at its head. The cloud is associated with the larger complex of star-forming regions that are near it in the plane of the sky (Figures \ref{4402sefilament}, \ref{4402ereg}). Since Cloud 10 is in the transition zone, it is probably located at a larger galactocentric radius than the star-forming regions located to either side of it in projection. 

There are significantly more bright blue stars in the eastern section than in the other parts of the transition zone. Has recent star formation been triggered by ram pressure, or are young stars simply being revealed on the leading side of the galaxy as ram pressure strips away the dust surrounding star-forming regions? Although the UV emission from the southeast star-forming regions is very bright \citep{GildePaz07}, a map of ongoing star formation made by combining Spitzer IR and GALEX UV emission \citep{Vollmer12} suggests that the eastern region of the galaxy has a star formation rate comparable to other parts of the galactic disk that do not have as many visible bright stars and clusters. It is possible that star formation in the eastern region has been modestly enhanced by a factor of $\sim$2 relative to other parts of the disk at the same galactocentric radius. Pre-existing substructure likely plays a role in the different appearances of the two sides of the galaxy---the eastern, more star-forming side may be located in a spiral arm, and the western side may be an interarm region. 

The eastern side of the galaxy also hosts a large, extraplanar dust complex extending several kpc above the disk (outlined in blue in Fig. \ref{4402ereg}), surrounded by 9 isolated and semi-decoupled dust clouds. The complex extends up to 2.2 kpc north of the major axis, whereas in the rest of the galaxy, highly obscuring dust clouds only extend to $\sim$400 pc north of the major axis. Three of the small clouds are elongated, which is evidence that the ICM wind may be affecting them, but the substructures in this complex are not preferentially elongated in the projected wind direction. The monolithic nature of this complex is similar to the upturn structure in NGC 4522, although there is not a distinct ridge of young stars beyond the extraplanar dust lane in NGC 4402. We further discuss the similarities and differences between the two structures in Section \ref{simdiff}.

\subsubsection{Description of N4402 Western Dust Front Complex and Pointy Dust Knots}
\label{pdks}

The western side of NGC 4402 near Cloud 20 contains a long ($\sim$3--4 kpc), smooth, continuous ``dust front" (labeled in Figure \ref{4402wfilament}) that appears to be located in or near the disk plane, as well as a ``long northern dust ridge" which is probably extraplanar. No smaller, decoupled clouds are detected south of the dust front. The ridge is significantly less extincting than Cloud 20 and appears to cross behind it. We suspect that the dust front is actually located in the disk plane, while the part of Cloud 20 that overlaps it is extraplanar, since it is closer to us than the rest of the structure. The head of Cloud 20 could be in or near the disk plane at a larger galactocentric radius. The western side of the galaxy also contains very few young stars, in contrast with the eastern region's widespread star formation. 

We also note several strongly-extincting dust features that are located northwest of Cloud 20 and the dust front, labeled as ``pointy dust knots" in Figure \ref{4402wfilament}, which are part of the larger ``long northern cloud complex" of highly extincting clouds. The three pointy dust knots have v-shaped morphologies, and the vertex of the ``v" at the southern end of each cloud is significantly extincting with a sharp southern boundary. The central one has a more curved v shape resembling a tornado. To the north of each v, the clouds become somewhat more diffuse. The easternmost pointy dust knot has the highest extinction to the southeast, with a gradual decrease in extinction toward the northwest, suggesting a head-tail structure. It is also the largest, with a longest dimension of $\sim$350 pc. The other two have more uniform surface brightnesses and shorter lengths, with their longest dimension $\sim$100 pc. We suggest that the pointy dust knots are early in the process of decoupling from the surrounding ISM and have the potential to become fully decoupled clouds like the ones in the transition zone. Just west of the pointy dust knots, there are several other elongated structures that extend south of the same large, dark complex as the pointy dust knots. These elongated structures have less clearly-defined southern edges than the pointy dust knots and may be at an even earlier stage of decoupling. 

If the dust front to the south is located in the disk plane, then these v-shaped clouds and the rest of the long northern cloud complex must be above the disk plane. This is also supported by the fact that the v-shaped clouds are very dark, but are projected against the far side of the disk---if they actually were located on the far side of the galaxy, they would block less starlight along the line of sight, and would appear less dark. The pointy dust knots may be within a part of the ISM that has been pushed downwind and above the disk from its original location in the disk plane. The less-extincting smooth dust front to the south (black line in Figure \ref{4402wfilament}) may be at a smaller galactocentric radius that is more tightly gravitationally bound and thus less susceptible to ram pressure.

The extraplanar long northern cloud complex and pointy dust knot substructures in NGC 4402 suggest that the ICM in those regions is gradually pushing the ISM out of the disk, and the continuous fronts, ridges, and linear structures suggest that each parcel of ISM gas is influenced by the behavior of the gas to either side of it. As it is moving, substructure develops due to decoupling and/or instabilities. There is polarized radio continuum emission coincident with the entire complex of the dust front, Cloud 20, and the pointy dust knots (Figure \ref{4402rc}). The polarized radio continuum shows the presence of a large-scale, ordered magnetic field in the disk and/or halo like the one outside the NE dust upturn in NGC 4522 (Section \ref{upturnmorph}) and suggests that the ICM wind is not flowing freely through the disk plane in a way that would disrupt a large-scale magnetic field.

\subsubsection{Large Filament Comparison---Size \& Substructures}
\label{filamentcomp}
\begin{figure*}[htb] 
  \centering
  \includegraphics[width=5in]{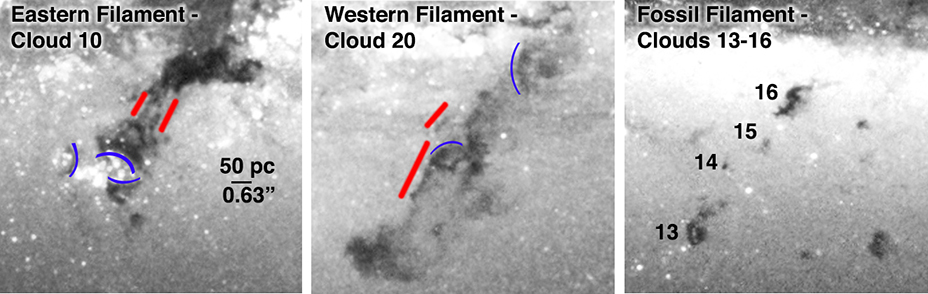} 
  \caption{NGC 4402 Clouds 10 (the eastern dust filament) and 20 (the western dust filament), and Clouds 13-16 (the possible fossil filament, which may consist of the densest remaining substructures after the less-dense areas have been stripped). Linear substructures have been highlighted with red lines, and arcs are shown in blue. The scale bar for the three panels is shown in the left panel.}
  \label{4402filaments}
\end{figure*}

In Figure \ref{4402filaments}, we compare the large NGC 4402 dust filaments, Clouds 10 and 20, in order to constrain how their ICM substructures have been affected by ram pressure stripping. The head and middle portion of Cloud 10 align approximately with the projected ICM wind angle, but the base is at a $\sim120\degree$ angle to the middle segment of the cloud. The base connects to the galaxy's main dust distribution and appears as a highly obscuring cloud in Figure \ref{4402filaments}. The head, which is less extincting, has significant substructure, some of which is indicated with red and blue lines. There are also two parallel linear substructures in the less-dense middle segment of the cloud that align with the head and middle's position angle and the projected wind direction. The substructures are on the boundary between the head and base (red), which may be the result of the ICM wind stretching out part of the cloud. The head contains knots and filaments with a characteristic spatial scale of $\sim$20--80 pc. In the head of the cloud, there are at least four elongated clouds forming arcs (Figure \ref{4402filaments}, marked in blue) that are associated with the region of bright, young stars at the head of the filament. The substructure in the head of Cloud 10 appears to be influenced more by star formation than by ram pressure. 

\begin{figure*}[htb] 
  \centering
  \includegraphics[width=5in]{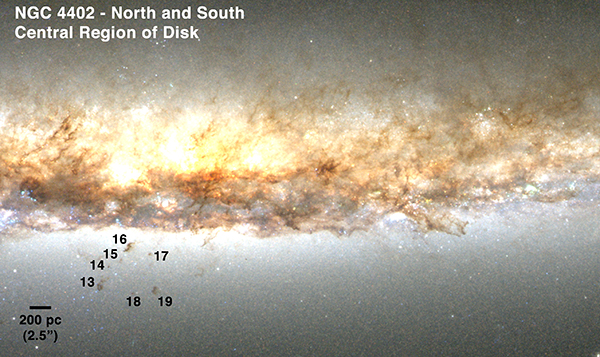} 
  \caption{NGC 4402 central dust region---the area south of the disk is largely free of small dust clouds, with only one group of them. The possible fossil filament, described in Section \ref{filamentcomp}, is made up of clouds 13--16.}
  \label{4402sreg}
\end{figure*}

Cloud 20 contains some dense, complex substructure, but it also contains more tenuous, diffuse dust than Cloud 10 (Figure \ref{4402filaments}). Notably, Cloud 20's tail is not centered behind its head, but is offset somewhat toward the west. Substructure throughout the cloud is not uniform, with relatively little substructure in the head and a number of dense regions further up. In general the cloud is more extincting on the eastern side, with a relatively well-defined eastern boundary, and becomes gradually less dark toward the west side. Like Cloud 10, Cloud 20 also has some linear substructures, located on the eastern edge above the head (marked with red lines in Figure \ref{4402filaments}). There is not a smooth gradient from east to west or from the head to the tail---the linear substructures border a diffuse area about 100 pc long with little structure. Near the midpoint of the cloud's length, substructure is visible in clumps that are are on average somewhat larger than the ones in Cloud 10, with a typical longest dimension of $\sim$60 pc. There are only two arc-like structures (Figure \ref{4402filaments}, blue). The northern few hundred pc of the cloud contains some diffuse areas and some substructure that is not as obscuring as the substructure peaks at the middle of the cloud.  
 
The differences in size and star formation between Cloud 10 and Cloud 20 may be attributable to a difference in ages, but the two are probably not simply at different evolutionary stages---pre-existing substructure in the material the clouds formed from is likely a major factor. Some differences may be attributable to the physical processes that formed the clouds. Since Cloud 20 is longer and less dense but has a similar mass to Cloud 10, it may be dynamically older---the wind may have been working to elongate it for a longer time. The head of Cloud 20 is not forming stars, while the head of Cloud 10 did. If the two were simply at different evolutionary stages, we would probably see at least some evidence of early-stage star formation in Cloud 20, which we do not. Apparently, a substructure clump like the head of Cloud 20 can be dense enough to decouple from the rest of the ISM during stripping even if it is not dense enough to form stars. The two clouds have sufficiently different structure that we cannot tell which is older simply by the size or amount of star formation. 

It is possible that Clouds 13-16, which are grouped together at $\sim$1--1.5 kpc (12$\arcsec$--18$\arcsec$) southeast of the galactic center (Figure \ref{4402sreg}), form a ``fossil filament"---if the four clouds were originally part of a larger elongated complex, with less-dense material in between them, then it is possible that in the face of sufficient pressure over time, the less-dense material has been stripped away, leaving only the densest substructures behind in a fossil filament. The clouds form a structure that is aligned along an axis with a similar position angle to the two large filaments (330$\degree \pm 10 \degree$; \citealt{Abramson14}), and the clouds are roughly the same size as the substructure clumps in clouds 10 and 20. In the fossil filament scenario, the aligned clouds represents a later evolutionary stage than Clouds 10 and 20. 

\subsubsection{Elongated Dust Filament and Front Interpretation---Magnetic Fields}
\label{magfields}

The existence of significant substructure in the heads and tails of both kpc-scale dust filaments in NGC 4402 (Clouds 10 and 20, Figure \ref{4402filaments}) appears to conflict with the interpretation advanced by \citet{crowl05} of how they formed. The two filaments were first identified by \citet{crowl05} in ground-based images with 0$\arcsec$.5 (40 pc) resolution. The clouds' position angles lead \citet{crowl05} to speculate that they are being ablated by the ICM wind, with the outer layers of the cloud being stripped off and swept into a tail that trails behind the denser head of the cloud. The observational signature of ablation would be a highly-extincting head, possibly with a bow shock, with a more-diffuse, relatively smooth tail trailing behind it, perhaps being shielded from the ICM wind by the cloud's head. The cloud might have a v-shaped morphology with an opening angle behind the head, although depending on the density structure and our detection limits, the structure might still appear linear. The exact structure of the tail would depend on factors such as the wind speed, duration of stripping, and the original structure of the cloud (e.g., spherical or fractal, \citealt{cooper09}). 

Now that we have observed the significant substructure and highly-extincting tails in Clouds 10 and 20, an alternative explanation emerges of their morphology---they are clearly more complex than simple homogeneous clouds being ablated from the outside in. Several different phenomena could be at work to create the complex structures we see. Shadowing probably plays some role---when the disk is stripped except for the densest clouds, i.e., the upwind ends of the filaments, a dense cloud blocks the wind from stripping the gas in its wake, leaving a column-shaped dust cloud containing density variations similar to what they were before the intervention of the ICM wind. The clouds' pre-existing substructure means that they will not have the smooth sides characteristic of simple ablation from the outside in. Neither cloud is completely linear---the tail of Cloud 20 is not centered behind the head but is offset to the west, and the head and middle of Cloud 10 adjoin smoothly to its highly extincting base, which is at an angle to the head and appears to be connected to the galaxy's main dust lane. The base of Cloud 10 is probably a sufficiently dense substructure to withstand the ICM wind. 

Shadowing and pre-existing substructure could produce some of the features we describe, but other features are difficult to explain with those mechanisms. The linear substructures at the edges of Clouds 10 and 20, marked on Figure \ref{4402filaments}, are also elongated roughly in the projected wind direction. There are two 50 pc long linear filaments in Cloud 10 that are parallel to each other and the projected wind direction. There is a longer, thinner filamentary substructure in Cloud 20 with a length of $\sim$200 pc and a width of 15 pc, which has possibly been stretched more than the Cloud 10 substructures by the ICM wind. In addition to the elongated substructures, we note the wide range of dust densities in the two large filaments. Although the denser substructures are of similar size and density to the smaller, isolated, decoupled transition zone clouds elsewhere in the galaxy, Cloud 10 and especially Cloud 20 also contain regions that are much more diffuse than any isolated clouds are---presumably, gas of a similar density everywhere in the transition zone that isn't anchored by a large density peak (like the filaments' heads) has already been stripped. The presence of elongated substructures and regions that would have been stripped without the surrounding denser clumps suggests that the cloud as a whole may be partially bound together. 

Given the evidence for magnetic field binding during ram pressure stripping in NGC 4921 (discussed further in Section \ref{magbinding}), we consider a magnetic field explanation for the formation of elongated, multi-density structures in NGC 4402. In this scenario, the linear substructures formed along with the elongated clouds as a whole, since they have the same position angle, and the clouds are partially decoupled, magnetically connected filaments whose heads are clouds too dense to strip directly. Frozen-in magnetic fields could have bound together dense substructures with less-dense ISM, while the ICM wind elongated the whole structure in the projected wind direction. We note that the size of the substructure clumps in Clouds 10 and 20, as well as many of the clumps in the galaxy's main dust lane, is comparable to the size of the isolated dust clouds in both galaxies, which is consistent with the clumps existing before the linear structure formed. We think that some of the current substructure is the result of ram pressure stretching out magnetically linked clouds, such as the linear substructures, and some of it is the result of pre-existing clumps of material remaining relatively intact as the filament forms. 

In this scenario, the dusty substructures that make up the large filaments are bound to each other by magnetic fields, and as the tail of accelerated gas and dust recedes from the densest clouds, the less-dense magnetically bound gas stretches. Longer filaments may be older, since the ICM wind should stretch the filament progressively more over time. However, the filaments are unlikely to become perfectly aligned with the wind direction, since they are probably anchored by several dense substructures. The offset between the head and tail of Cloud 20 can be explained in this scenario. Some of the substructure in Cloud 10 is also due to star formation and star formation feedback, especially from the prominent star-forming region in the cloud's head. Although many different processes are responsible for forming the clouds' substructure, magnetic binding of denser and less-dense areas during the stripping process may have helped to shape these elongated clouds. 

It is unclear whether the smaller elongated clouds in the two galaxies are the result of ablation or not. All have at least some substructure, though some have morphologies that may be consistent with ablation. We do not have sufficient resolution to explore their origin in detail. If the smaller elongated clouds have not been ablated, it means that their formation, like that of the larger elongated clouds, depends on magnetic fields connecting together a small parcel of ISM as it stretches out along the wind direction. 

\subsubsection{Magnetic Binding in Simulations and Other Galaxies}
\label{magbinding}

Relevant simulations suggest that magnetic fields are useful in explaining the elongated clouds' morphologies. Using simulations of ram pressure stripping that feature a galactic magnetic field, \citet{Tonnesen14} find that inclusion of magnetic fields suppresses mixing in the stripped gas. As a consequence, more intermediate-density substructures stay intact during stripping and persist in the tail, although the densest gas still is not directly stripped from the disk. This result is broadly consistent with our scenario in which magnetic fields play a role in keeping dense clouds together during the stripping process. In simulations that feature magnetic fields only in the ICM, as well as radiative cooling, filamentary structures are produced in the stripped gas tail during ram pressure stripping \citep{Ruszkowski14}. Other groups have simulated the interaction of a supersonic wind with a dense cloud in the context of galactic winds and supernova remnants. It is thought that the presence of magnetic fields may suppress thermal conduction at the cloud's boundary layer and protect somewhat against the Kelvin-Helmholtz instabilities that are thought to be ultimately responsible for breaking up clouds in a streaming plasma \citep[e.g.,][]{Orlando08, Yamagami11}. 

Observationally, the elongated clouds and pointy dust knots in NGC 4522 and NGC 4402 bear a morphological resemblance to the ``elephant trunks" of \citet{carlqvist98, carlqvist02, carlqvist03}, elongated clouds with significant substructure found in Milky Way star-forming regions. \citet{carlqvist03} advance a theory in which pressure from the expanding gas of an HII region will cause the ISM of varying densities bound by a magnetic field to assume a v shape, where the vertex of the v is a region of denser ISM with enough inertia to resist being pushed by the flow of ionized gas longer than the more diffuse gas that moves downwind to form the legs of the v. Carlqvist has noted similar, larger-scale ``mammoth trunks" in the galaxies NGC 1316 \citep{carlqvist10} and NGC 4921 \citep{carlqvist13} that may also result from magnetized ISM becoming elongated. However, he does not identify the force that makes these filaments.
\citet{kenney15} make the case that external ICM ram pressure is the causative agent in NGC 4921. The cause in NGC 1316 is different, probably ram pressure originating within the galaxy from an AGN outburst.


\begin{figure*}[htb] 
  \centering
  \includegraphics[width=6in]{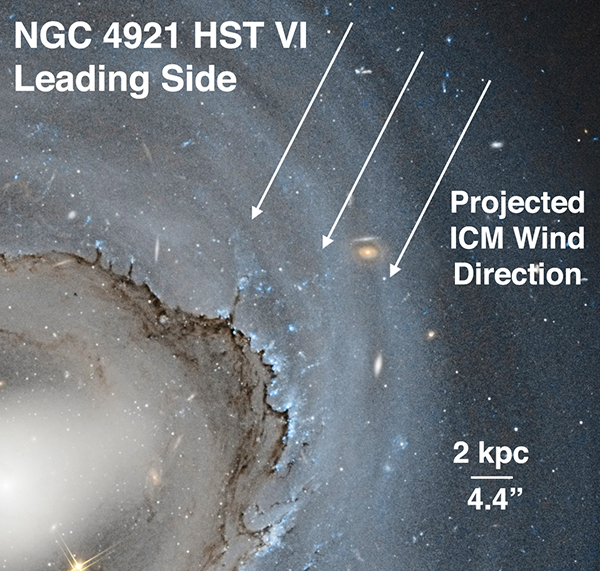} 
  \caption{HST V- and I-band image (with pseudogreen) of the leading side of NGC 4921, an anemic spiral galaxy in the Coma Cluster that shows features similar to NGC 4402 including large elongated clouds and long, smooth dust fronts. Image from R. Colombari.}
  \label{n4921}
\end{figure*}

\begin{figure*}[htb] 
  \centering
  \includegraphics[width=5in]{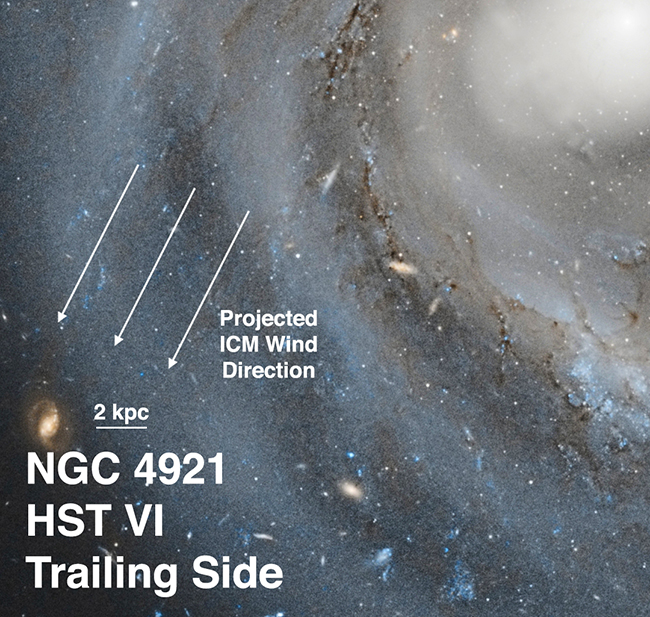} 
  \caption{HST V- and I-band image (with pseudogreen) of the trailing side of NGC 4921. The trailing side shows no elongated clouds or dust fronts, only small substructures without common position angles. Image from R. Colombari.}
  \label{n4921trail}
\end{figure*}

\section{Similarities and Differences Between the Two Galaxies and NGC 4921, A Stripped Coma Spiral}
\label{simdiff}

Comparing multiple galaxies gives us a fuller picture of what happens during stripping. In Figures 
\ref{4522rpsfeatures} and \ref{4402rpsfeatures}, we have summarized the structures described in this paper that are likely related to ram pressure stripping. We think that most of the observed differences between the ISM structures in NGC 4522 and NGC 4402 are related to differences in evolutionary stage of stripping, ram pressure strength, pre-existing substructure, wind angle, and viewing angle. We can use an example of a face-on stripped galaxy, in which we can see both the leading and trailing sides of the stripping interaction, to constrain which structures in each galaxy may be characteristic of a given viewing angle. By comparing with face-on stripped Coma galaxy NGC 4921 (this paper Figure \ref{n4921}; \citealt{kenney15}), we can make a case that certain features are generally present only on the leading or trailing side of a stripped galaxy. NGC 4921 has a well-defined continuous dust front on its leading side that is over 20 kpc long and extends azimuthally through 90$\degree$ of the galaxy (Figure \ref{n4921}). The dust front has significant substructure---a number of filaments elongated approximatly in the wind direction connect to it, as well as v-shaped clouds with their vertices located on the upwind sides. The trailing side, however, has no large dust fronts and few structures elongated in the projected ICM wind direction (Figure \ref{n4921trail}).

The similarity between the structures in NGC 4402 and NGC 4921 and the lack of clearly analogous kpc-scale filaments and well-defined dust fronts in NGC 4522 is likely due to our viewing angle. In dust extinction, we see the leading side of NGC 4402 and the trailing side of NGC 4522. Comparison with NGC 4921 (Figure \ref{n4921}) shows that dust filaments and fronts may form preferentially on the leading side of a galaxy experiencing ram pressure. 

 \begin{figure*}[htb] 
  \centering
  \includegraphics[width=6in]{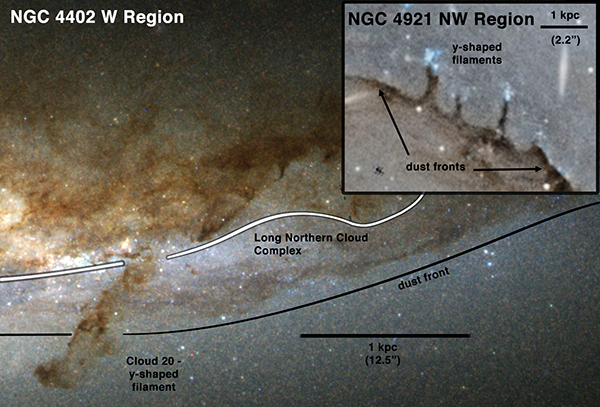} 
  \caption{NGC 4402 western region with inset of similar structures in face-on Coma spiral NGC 4921 \citep{kenney15}. The dust fronts in NGC 4921 have similarities with both the long northern cloud complex and the dust front in NGC 4402. All of these regions have a low density of young stars compared to other parts of these two galaxies.}
  \label{n4402-w-4921}
\end{figure*}

 \begin{figure*}[htb] 
  \centering
  \includegraphics[width=6in]{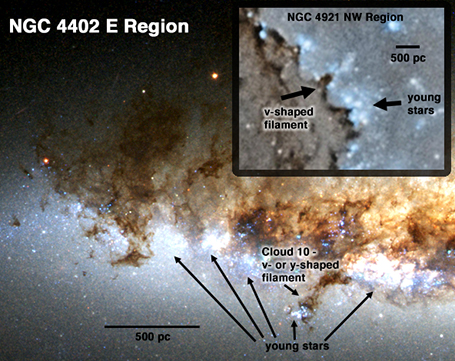} 
  \caption{NGC 4402 eastern region with inset of similar structures in face-on Coma spiral NGC 4921. Both regions have a high density of young stars.}
  \label{n4402-e-4921}
\end{figure*}

The NGC 4402 western filament, Cloud 20 (Figure \ref{n4402-w-4921}) and the pointy dust knots seem to extend from a long, continuous dust front that includes the long northern cloud complex. The pointy dust knots and associated cloud complex could also form a dust front when viewed from above, and along with the western filament, they may be substructure within the larger dust front. Cloud 20 has a similar extinction to the long northern cloud complex, indicating that they are probably at a similar depth in the galaxy and may be physically associated with each other like the filaments and dust front in NGC 4921. Like Cloud 20, the longest NGC 4921 filaments are kpc-scale in projected length, although the filaments' total 3D length may be longer since there is probably a component along the line of sight. 

It appears that the smooth dust front in NGC 4921 is created by the ICM pushing the ISM further back into the disk, while dense clouds partially decouple to create the linear filaments and v-shaped clouds. The clouds only partially decouple (at least initially), leading us to believe that something is binding together denser and less-dense gas in the clouds and within the dust front itself. There is much substructure in the ISM, so without a force, possibly magnetic, holding together gas of different densities, each part of the dust front would be accelerated at different rates depending on the local gas density, and the front would not be as smooth. 

We hypothesize that the structures along the leading side of NGC 4921 are at different evolutionary stages of decoupling, which is discussed in detail by \citet{kenney15}. We think that initially, an overdensity partially decouples to create a v-shaped structure similar to the pointy dust knots in NGC 4402 (Figure \ref{4402wfilament}), and over time as the lower density dust front is pushed further back into the disk, the head of the structure elongates and the legs of the v move closer together. Eventually, kpc-scale filaments with y-shaped morphologies are formed, which consist of a distinct, narrow head, and connect to the large-scale dust front by filaments that angle outwards from the head, sometimes asymmetrically. The head and tail parts of the filament continue to differentiate, with the head becoming straighter, and the forked part of the "y" near the base remaining curved. 

NGC 4402 cloud 10 seems to fall along this evolutionary sequence (Figure \ref{n4402-e-4921}). If Cloud 10 were viewed from above, like the structures in NGC 4921, it would probably look very similar to the one labeled ``v- or y-shaped filament" It is shorter than Cloud 20 (2/3 the length) and that may be because it is at an earlier evolutionary stage and has not yet been stretched out as far by the ICM wind. Like Cloud 20, Cloud 10 appears to be connected to the galaxy's main dust lane. The northern part of this filament, at its base where it adjoins to the main dust lane, has similar extinction and substructure to the large, highly-extincting dust complex above it in the disk, indicating that they are probably physically associated with each other. The base, which is at an angle to the head and southern part of the cloud, may be part of the main ISM front. Based on NGC 4402 and NGC 4921, we have good evidence that kpc-scale filaments can form in the approximate wind direction at the leading edges of galaxies experiencing an ICM wind. 

Ram pressure strength is probably an important cause of differences between the three galaxies. It is certainly a factor in the differences between NGC 4921 and the Virgo galaxies, since ram pressure in Coma is generally 1--2 orders of magnitude stronger as a result of an average orbital velocity $\sim$3 times higher than that of Virgo and an ICM with peak density an order of magnitude higher. Ram pressure strength also differs significantly among galaxies in any cluster, which may help explain the differing structures in the two Virgo galaxies. A key difference between the two Virgo galaxies is that there is significant extraplanar star formation in NGC 4522, one example of which is the extraplanar arm (Section \ref{xparm}), and not very much in NGC 4402. This could be because NGC 4522 has experienced more recent strong stripping, the evidence for which is discussed in Section \ref{summaryofwork}. 

The difference in current ram pressure strength may also help explain the difference between the stellar and dust upturns in NGC 4522 and their possible counterpart in NGC 4402, the large eastern extraplanar dust plume (Section \ref{4402east}) that extends $\sim$2.2 kpc north of the major axis. It is not an exact counterpart to the NGC 4522 dust plume; the edge of the structure in NGC 4402 is less defined, and there are significantly more transition zone clouds just outside of it. While some of the decoupled clouds around it are elongated, the dust plume itself has no visible elongated substructures. The differences might be explained if both galaxies are experiencing ram pressure, but NGC 4522 is currently experiencing stronger pressure. As a result, the upturn structure on the leading side of NGC 4522 is more well-defined.

Our viewing angle with respect to the leading side may be the reason we observe more decoupled clouds in the transition zone (within $\pm$1 kpc from the disk plane) in NGC 4402 than in NGC 4522---there are only two decoupled clouds in NGC 4522 within 1 kpc of the disk plane, while 17 of the clouds in NGC 4402 qualify. As noted in Paper 1, stripping and decoupling may happen preferentially in the leading half of the galaxy, so we view more recently decoupled clouds in NGC 4402. The clouds we see in NGC 4522 are up to half a rotational period away from the leading side where they may have decoupled, so additional decoupled clouds could have been destroyed during the $\sim$200 Myr it took for the leading-side clouds to rotate around to the area where we can see them. 

The effects of the ICM wind are probably stronger on the trailing side of NGC 4522 than they are in NGC 4921 because the disk-wind angle in 4921 is probably closer to edge-on in spite of a significant line-of-sight component to its velocity \citep{kenney14}, so the trailing side is fairly shielded from the direct effects of the ICM wind. The disk-wind angle in NGC 4522 is closer to face-on (60$\degree \pm 10 \degree$;\citealt{Vollmer06}), so the trailing side will be less shielded, and therefore more strongly stripped. The projected disk-wind position angle in NGC 4402 is $\sim$45$\degree$ \citep{crowl05}, but there is also a significant component along the line of sight, since the galaxy is blue shifted with respect to the cluster velocity by almost 800 km s$^{-1}$. The 3D disk-wind angle in NGC 4402 is $\leq$45$\degree$ according to the technique used to constrain wind angles by \citet{Abramson11}. Since we view the leading side of NGC 4402, shielding is not as relevant to our interpretation of its characteristics, but the disk-wind angle is likely responsible for the position angle of the large filaments.  

\section{Conclusions}
\label{conclusions}

In Virgo spirals NGC 4522 and NGC 4402, we observe a number of distinct ISM dust features that strongly constrain the physical processes that occur during an ICM-ISM ram pressure interaction. For an overview of many of the features discussed below, see Figures \ref{4522rpsfeatures} (NGC 4522) and \ref{4402rpsfeatures} (NGC 4402).
\begin{enumerate}

\item We observe a striking upturn structure in the NE region of NGC 4522 at the edge of the remaining disk ISM (Figure \ref{4522nereg}), consisting of a dust upturn curving up and out of the disk and a young stellar upturn located $\sim$1 kpc radially beyond the dust upturn. The dust upturn has two distinct parts: a moderately sloping dust lane at smaller galactocentric radii and a steeply sloping dust plume at larger radii. Below the dust upturn, the only dust extinction in the disk plane is one isolated transition zone cloud. The stellar upturn curves out of the disk at a similar angle to the dust upturn and has many young, blue stars, but only one area with ISM and ongoing star formation (Figure \ref{4522-upturn}), which is located at the top of the upturn $\sim$2 kpc from the disk plane. There is a ridge of polarized radio continuum emission roughly coincident with the bottom $\sim$1 kpc of the stellar upturn (Figure \ref{4522rc}). 

We propose that the dust upturn is made up of ISM that is in the process of being stripped from the disk by ram pressure, with kpc-sized regions of the ISM moving coherently. In the past, dense clouds collapsed from stripped ISM and formed the stars now present in the stellar upturn. The stellar and dust upturns may be different evolutionary phases of the same phenomenon as the galaxy is stripped from the outside in.

\item In the SW extraplanar region of NGC 4522 (Figure \ref{4522swreg}), we observe a significant number of young, blue stars, including $\sim$100 pc-scale groupings extending up to 3.4 kpc above the major axis. The groupings begin near the minor axis and extend to a radius of $\sim$5 kpc. Kinematically distinct HI is associated with the distribution of blue stars (Figure \ref{4522swhiha}). A number of dense, decoupled dust clouds are present 500 pc---1 kpc above the otherwise-stripped disk plane and below the young stars (Figure \ref{4522sw-wind}).
 
We propose that the extraplanar SF regions are part of a distinct extraplanar arm or arm-like structure formed during stripping, as proposed by the models of Vollmer et al. (2008). The dense, decoupled clouds between the disk plane and the extraplanar arm may have decoupled from the material that was stripped and subsequently formed the extraplanar arm. 

\item On the western side of NGC 4402, we observe several distinctive elongated dust structures (Figure \ref{4402wfilament}). The southern (upwind) boundary of the dust distribution is marked by a 3--4 kpc long, smooth dust front oriented roughly perpendicular to the projected wind direction. Above the dust front, and likely above the disk plane and at a larger galactocentric radius, is the long northern cloud complex, which is more highly obscuring and shows more substructure, including pointy dust knots. A ridge of polarized radio continuum is coincident with the dust front and pointy dust knots. 

We propose that the dust front is formed when relatively low-density gas is pushed up together by the ICM wind. The long northern cloud complex also indicates an area in which the ICM is pushing most of the ISM up out of the disk at once, and the radio continuum ridge indicates the presence of large-scale ordered magnetic fields near the ICM-ISM interaction boundary (Figure \ref{4402rc}). We suggest that magnetic fields are needed to bind gas together in order to make relatively straight, smooth fronts such as the dust front and the long northern cloud complex. Without magnetic fields, each parcel of gas along the front would be accelerated at a different rate according to the local densities, which can vary significantly, and the front would not be so smooth. Magnetic fields also suppress turbulent mixing that could otherwise disrupt a smooth dust front.

\item NGC 4402's transition zone, a partially-stripped region of the disk just outside the main dust truncation region, contains kpc-scale dust filaments first described by Crowl et al. (2005) (Figure \ref{4402filaments}). In these filaments, we observe substructure on the same spatial scales (30--150 kpc) as the small clouds in the transition zones. This substructure includes linear substructures parallel to the filament direction with spatial scales of 50--100 pc. 

The kpc-scale filaments lack a counterpart in NGC 4522, which may be because large elongated filaments only form on the leading side of a galaxy experiencing ram pressure (e.g., the leading side of the face-on galaxy NGC 4921, Figure \ref{n4921}). Due to our viewing angle, in dust extinction we see the leading side of NGC 4402 and the trailing side of NGC 4522. It is possible that after further exposure to ram pressure, the less-dense material in the filaments will be stripped away, leaving them with a morphology like the Cloud 13-16 fossil filament.

We suggest that the large filaments are the result of ram pressure acting on massive dense clouds that are bound to the surrounding multi-density ISM by magnetic fields. Like the dust front, the large filaments may be an example of magnetic fields binding together ISM of varying densities to form one elongated, filamentary structure. The long, continuous dust fronts and elongated filament structures in NGC 4402 are morphologically similar to those in NGC 4921 (Figures \ref{n4402-w-4921} and \ref{n4402-e-4921}). 

\item We observe dense clouds on a variety of spatial scales that have decoupled or are decoupling from the surrounding less-dense ISM. Both galaxies have decoupled dense dust clouds outside their main dust truncation radii (Clouds 13 and 19 in NGC 4522 and NGC 4402, respectively), including the kpc-scale filaments in NGC 4402 first described by Crowl et al and $\sim$10--100 kpc-scale isolated clouds. It appears that some of the densest gas remains in or near the disk plane in spite of ram pressure, while the less-dense ISM around it is stripped away.

We propose that the pointy dust knots in NGC 4402 (Figure \ref{4402wfilament}) may be partially-decoupled, magnetically-bound clouds like the NGC 4402 kpc-scale filaments (Figure \ref{4402filaments}). At early stages of stripping, dense clouds may partially decouple, creating v-shaped filaments like the pointy dust knots. Over time, or if the initial cloud was fairly isolated, a more linear head-tail filament may form after further exposure to the ICM wind.    

\end{enumerate}

Detailed observations of the ISM during stripping show that the structures created during an ICM-ISM ram pressure interaction are strongly influenced by pre-existing ISM substructures, the ICM wind angle and strength, the duration of ram pressure, and possibly the presence of magnetic fields to bind together denser and less-dense gas.

 \bibliography{hstrefs2}

\end{document}